\documentclass[copyright,creativecommons]{eptcs}

\providecommand{\event}{GANDALF 2023}

\usepackage[utf8]{inputenc}
\usepackage{bold-extra}
\usepackage{amsmath,
	    amsfonts,
	    amstext,
	    amssymb,
	    mathrsfs,
	    mathpartir,
	    stmaryrd,
	    latexsym,
	    enumitem,
	    proof,
	    graphicx,
	    color,
	    hyperref,
	    comment,
	    semantic,
	    float,
	    hhline,
	    tikz}

\newcommand{\ms}[1]
	{\null\ifmmode\mathord{\mathcode`-="702D\it #1\mathcode`\-="2200}
	\else$\mathord{\mathcode`-="702D\it #1\mathcode`\-="2200}$\fi}

\newcommand{\cws}[2]
	{\\ \centerline{$#2$} \\[-#1pt]}

\newcommand{\fullbox}
	{{\mbox{}\nolinebreak\hfill{$\rule{2mm}{2mm}$}}}

\newcommand{\bibtrick}[1]
	{}

\newcommand{\calb}
        {\mathcal{B}}

\newcommand{\call}
        {\mathcal{L}}

\newcommand{\natns}
	{\mathbb{N}}

\newcommand{\procs}
	{\mathbb{P}}

\newcommand{\pt}
	{\textit{path}}

\newcommand{\initial}
	{\textit{initial}}

\newcommand{\final}
	{\textit{final}}

\newcommand{\reachable}
	{\textit{reachable}}

\newcommand{\arrow}[2]
        {\, {\auxarrow\limits^{#1}}_{#2} \,}
\newcommand{\auxarrow}
	{\mathop{\longrightarrow}}

\newcommand{\warrow}[2]
        {\, {\wauxarrow\limits^{#1}}_{#2} \,}
\newcommand{\wauxarrow}
	{\mathop{\Longrightarrow}}

\newcommand{\rarrow}[2]
        {\, {\rauxarrow\limits^{#1}}_{#2} \,}
\newcommand{\rauxarrow}
	{\mathop{- \!\!\! \rightsquigarrow}}

\newcommand{\nil}
	{\underline 0}

\newcommand{\sbis}[1]
	{\sim_{#1}}

\newcommand{\wbis}[1]
        {\approx_{#1}}

\newcommand{\diam}[2]
	{\langle #1 \rangle_{#2}}

\newcommand{\wdiam}[2]
	{\langle \! \langle #1 \rangle \! \rangle_{#2}}

\newcommand{\depth}
	{\textit{depth}}

\newtheorem{new_theorem}
	{Theorem}[section]

\newtheorem{new_definition}
	[new_theorem]{Definition}

\newtheorem{new_remark}
	[new_theorem]{Remark}

\newtheorem{new_example}
	[new_theorem]{Example}

\newtheorem{new_lemma}
	[new_theorem]{Lemma}

\newtheorem{new_proposition}
	[new_theorem]{Proposition}

\newtheorem{new_corollary}
	[new_theorem]{Corollary}

\newenvironment{definition}
	{\begin{new_definition}\rm}
	{\end{new_definition}}

\newenvironment{example}
	{\begin{new_example}\rm}
	{\end{new_example}}

\newenvironment{lemma}
	{\begin{new_lemma}\rm}
	{\end{new_lemma}}

\newenvironment{theorem}
	{\begin{new_theorem}\rm}
	{\end{new_theorem}}

\newenvironment{proof}
	{\medskip\noindent{\bf Proof}$\ $}
	{}

\begin{document}

\title{Modal Logic Characterizations of \\
       Forward, Reverse, and Forward-Reverse Bisimilarities}
\def\titlerunning{Modal Logic Characterizations of Forward, Reverse, and Forward-Reverse Bisimilarities}

\author{Marco Bernardo \qquad Andrea Esposito
\institute{Dipartimento di Scienze Pure e Applicate, Universit\`a di Urbino, Urbino, Italy}}
\def\authorrunning{M.~Bernardo, A.~Esposito}

\maketitle


\begin{abstract}
Reversible systems feature both forward computations and backward computations, where the latter undo the
effects of the former in a causally consistent manner. The compositionality properties and equational
characterizations of strong and weak variants of forward-reverse bisimilarity as well as of its two
components, i.e., forward bisimilarity and reverse bisimilarity, have been investigated on a minimal process
calculus for nondeterministic reversible systems that are sequential, so as to be neutral with respect to
interleaving vs.\ truly concurrent semantics of parallel composition. In this paper we provide logical
characterizations for the considered bisimilarities based on forward and backward modalities, which reveals
that strong and weak reverse bisimilarities respectively correspond to strong and weak reverse trace
equivalences. Moreover, we establish a clear connection between weak forward-reverse bisimilarity and
branching bisimilarity, so that the former inherits two further logical characterizations from the latter
over a specific class of processes.
\end{abstract}

%
%
\section{Introduction}
\label{sec:intro}
%
%

Reversibility in computing started to gain attention since the seminal works~\cite{Lan61,Ben73}, where it
was shown that reversible computations may achieve low levels of heat dissipation. Nowadays \emph{reversible
computing} has many applications ranging from computational biochemistry and parallel discrete-event
simulation to robotics, control theory, fault tolerant systems, and concurrent program debugging. 

In a reversible system, two directions of computation can be observed: a \emph{forward} one, coinciding with
the normal way of computing, and a \emph{backward} one, along which the effects of the forward one are
undone when needed in a \emph{causally consistent} way, i.e., by returning to a past consistent state. The
latter task is not easy to accomplish in a concurrent system, because the undo procedure necessarily starts
from the last performed action and this may not be unique. The usually adopted strategy is that an action
can be undone provided that all of its consequences, if any, have been undone beforehand~\cite{DK04}.

In the process algebra literature, two approaches have been developed to reverse computations based on
keeping track of past actions: the dynamic one of~\cite{DK04} and the static one of~\cite{PU07}, later shown
to be equivalent in terms of labeled transition systems isomorphism~\cite{LMM21}.

The former yields RCCS, a variant of CCS~\cite{Mil89a} that uses stack-based memories attached to processes
to record all the actions executed by those processes. A single transition relation is defined, while
actions are divided into forward and backward resulting in forward and backward transitions. This approach
is suitable when the operational semantics is given in terms of reduction semantics, like in the case of
very expressive calculi as well as programming languages.

In contrast, the latter proposes a general method, of which CCSK is a result, to reverse calculi, relying on
the idea of retaining within the process syntax all executed actions, which are suitably decorated, and all
dynamic operators, which are thus made static. A forward transition relation and a backward transition
relation are separately defined, which are labeled with actions extended with communication keys so as to
remember who synchronized with whom when going backward. This approach is very handy when it comes to deal
with labeled transition systems and basic process calculi.

In~\cite{PU07} \emph{forward-reverse bisimilarity} was introduced too. Unlike standard forward-only
bisimilarity~\cite{Par81,Mil89a}, it is truly concurrent as it does not satisfy the expansion law of
parallel composition into a choice among all possible action sequencings. The interleaving view can be
restored in a reversible setting by employing \emph{back-and-forth bisimilarity}~\cite{DMV90}. This is
defined on computation paths instead of states, thus preserving not only causality but also history as
backward moves are constrained to take place along the path followed when going forward even in the presence
of concurrency. In the latter setting, a single transition relation is considered, which is viewed as
bidirectional, and in the bisimulation game the distinction between going forward or backward is made by
matching outgoing or incoming transitions of the considered processes, respectively.

In~\cite{BR23} forward-reverse bisimilarity and its two components, i.e., forward bisimilarity and reverse
bisimilarity, have been investigated in terms of compositionality properties and equational
characterizations, both for nondeterministic processes and Markovian processes. In order to remain neutral
with respect to interleaving view vs.\ true concurrency, the study has been conducted over a sequential
processes calculus, in which parallel composition is not admitted so that not even the communication keys
of~\cite{PU07} are needed. Furthermore, like in~\cite{DMV90} a single transition relation has been defined
and the distinction between outgoing and incoming transitions has been exploited in the bisimulation game.
In~\cite{BE23} the investigation of compositionality and axiomatizations has been extended to weak variants
of forward, reverse, and forward-reverse bisimilarities, i.e., variants that are capable of abstracting from
unobservable actions, in the case of nondeterministic processes only.

In this paper we address the logical characterization of the aforementioned strong and weak bisimilarities
over nondeterministic reversibile sequential processes. The objective is to single out suitable modal logics
that induce equivalences that turn out to be alternative characterizations of the considered bisimilarities,
so that two processes are bisimilar iff they satisfy the same set of formulas of the corresponding logic.
Starting from Hennessy-Milner logic~\cite{HM85}, which includes forward modalities whereby it is possible to
characterize the standard forward-only strong and weak bisimilarities of~\cite{Mil89a}, the idea is to add
backward modalities in the spirit of~\cite{DMV90} so as to be able to characterize reverse and
forward-reverse strong and weak bisimilarities. Unlike~\cite{DMV90}, where back-and-forth bisimilarities as
well as modality interpretations are defined over computation paths, in our reversible setting both the
considered bisimilarities and the associated modal logic interpretations are defined over states.

Our study reveals that strong and weak reverse bisimilarities do not need conjunction in their logical
characterizations. In other words, they boil down to strong and weak reverse trace equivalences,
respectively. Moreover, recalling that branching bisimilarity~\cite{GW96} is known to coincide with weak
back-and-forth bisimilarity defined over computation paths~\cite{DMV90}, we show that branching bisimilarity
also coincides for a specific class of processes with our weak forward-reverse bisimilarity defined over
states. Based on the results in~\cite{DV95}, this opens the way to two further logical characterizations of
the latter in addition to the one based on forward and backward modalities. The first characterization
replaces the aforementioned modalities with an until operator, whilst the second one is given by the
temporal logic CTL$^{*}$ without the next operator.

The paper is organized as follows. In Section~\ref{sec:background} we recall syntax and semantics for the
considered calculus of nondeterministic reversible sequential processes as well as the strong forward,
reverse, and forward-reverse bisimilarities investigated in~\cite{BR23} and their weak counterparts examined
in~\cite{BE23}. In Section~\ref{sec:modal_logics} we provide the modal logic characterizations of all the
aforementioned bisimilarities based on forward and backward modalities interpreted over states. In
Section~\ref{sec:branching_bisim} we establish a clear connection between branching bisimilarity and our
weak forward-reverse bisimilarity defined over states. In Section~\ref{sec:concl} we conclude with final
remarks and directions for future work.

%
%
\section{Background}
\label{sec:background}
%
%

%
\subsection{Syntax of Nondeterministic Reversible Sequential Processes}
\label{sec:syntax}
%

Given a countable set $A$ of actions -- ranged over by $a, b, c$ -- including an unobservable action denoted
by~$\tau$, the syntax of reversible sequential processes is defined as follows~\cite{BR23}:
\cws{0}{P \:\: ::= \:\: \nil \mid a \, . \, P \mid a^{\dag} . \, P \mid P + P}
where:

	\begin{itemize}

\item $\nil$ is the terminated process.

\item $a \, . \, P$ is a process that can execute action $a$ and whose forward continuation is $P$.

\item $a^{\dag} \, . \, P$ is a process that executed action $a$ and whose forward continuation is inside
$P$.

\item $P_{1} + P_{2}$ expresses a nondeterministic choice between $P_{1}$ and $P_{2}$ as far as both of them
have not executed any action yet, otherwise only the one that was selected in the past can move.

	\end{itemize}

We syntactically characterize through suitable predicates three classes of processes generated by the
grammar above. Firstly, we have \emph{initial} processes, i.e., processes in which all the actions are
unexecuted:
\cws{0}{\begin{array}{rcl}
\initial(\nil) & & \\
\initial(a \, . \, P) & \Longleftarrow & \initial(P) \\
\initial(P_{1} + P_{2}) & \Longleftarrow & \initial(P_{1}) \land \initial(P_{2}) \\
\end{array}}
\indent
Secondly, we have \emph{final} processes, i.e., processes in which all the actions along a single path have
been executed:
\cws{0}{\begin{array}{rcl}
\final(\nil) & & \\
\final(a^{\dag} . \, P) & \Longleftarrow & \final(P) \\
\final(P_{1} + P_{2}) & \Longleftarrow & (\final(P_{1}) \land \initial(P_{2})) \, \lor \\
& & (\initial(P_{1}) \land \final(P_{2})) \\
\end{array}}
Multiple paths arise only in the presence of alternative compositions. At each occurrence of $+$, only the
subprocess chosen for execution can move, while the other one, although not selected, is kept as an initial
subprocess within the overall process to support reversibility.

Thirdly, we have the processes that are \emph{reachable} from an initial one, whose set we denote by
$\procs$:
\cws{8}{\begin{array}{rcl}
\reachable(\nil) & & \\
\reachable(a \, . \, P) & \Longleftarrow & \initial(P) \\
\reachable(a^{\dag} . \, P) & \Longleftarrow & \reachable(P) \\
\reachable(P_{1} + P_{2}) & \Longleftarrow & (\reachable(P_{1}) \land \initial(P_{2})) \, \lor \\
& & (\initial(P_{1}) \land \reachable(P_{2})) \\
\end{array}}

\noindent
It is worth noting that:

	\begin{itemize}

\item $\nil$ is the only process that is both initial and final as well as reachable.

\item Any initial or final process is reachable too.

\item $\procs$ also contains processes that are neither initial nor final, like e.g.\ $a^{\dag} . \, b \, .
\, \nil$.

\item The relative positions of already executed actions and actions to be executed matter; in particular,
an action of the former kind can never follow one of the latter kind. For instance, $a^{\dag} . \, b \, . \,
\nil \in \procs$ whereas $b \, . \, a^{\dag} . \, \nil \notin \procs$.

	\end{itemize}

%
\subsection{Operational Semantic Rules}
\label{sec:semantics}
%

According to the approach of~\cite{PU07}, dynamic operators such as action prefix and alternative
composition have to be made static by the semantics, so as to retain within the syntax all the information
needed to enable reversibility. For the sake of minimality, unlike~\cite{PU07} we do not generate two
distinct transition relations -- a forward one $\! \arrow{}{} \!$ and a backward one $\! \rarrow{}{} \!$ --
but a single transition relation, which we implicitly regard as being symmetric like in~\cite{DMV90} to
enforce the \emph{loop property}: every executed action can be undone and every undone action can be redone.

In our setting, a backward transition from $P'$ to $P$ ($P' \rarrow{a}{} P$) is subsumed by the
corresponding forward transition $t$ from $P$ to $P'$ ($P \arrow{a}{} P'$). As will become clear with the
definition of behavioral equivalences, like in~\cite{DMV90} when going forward we view $t$ as an
\emph{outgoing} transition of $P$, while when going backward we view $t$ as an \emph{incoming} transition of
$P'$. The semantic rules for $\! \arrow{}{} \! \subseteq \procs \times A \times \procs$ are defined in
Table~\ref{tab:semantics} and generate the labeled transition system $(\procs, A, \! \arrow{}{}
\!)$~\cite{BR23}.

	\begin{table}[t]

\[\begin{array}{|ll|}
\hline
\inferrule*[left=(Act$_{\rm f}$)]{\initial(P)}{a \, . \, P \arrow{a}{} a^{\dag} . \, P} &
\inferrule*[left=(Act$_{\rm p}$)]{P \arrow{b}{} P'}{a^{\dag} . \, P \arrow{b}{} a^{\dag} . \, P'} \\[0.3cm]
\inferrule*[left=(Cho$_{\rm l}$)]{P_{1} \arrow{a}{} P'_{1} \quad \initial(P_{2})}{P_{1} + P_{2} \arrow{a}{}
P'_{1} + P_{2}} \quad &
\inferrule*[left=(Cho$_{\rm r}$)]{P_{2} \arrow{a}{} P'_{2} \quad \initial(P_{1})}{P_{1} + P_{2} \arrow{a}{}
P_{1} + P'_{2}} \\
\hline
\end{array}\]

\caption{Operational semantic rules for reversible action prefix and choice}
\label{tab:semantics}

	\end{table}

The first rule for action prefix (\textsc{Act$_{\rm f}$} where f stands for forward) applies only if $P$ is
initial and retains the executed action in the target process of the generated forward transition by
decorating the action itself with $\dag$. The second rule for action prefix (\textsc{Act$_{\rm p}$} where p
stands for propagation) propagates actions executed by inner initial subprocesses.

In both rules for alternative composition (\textsc{Cho$_{\rm l}$} and \textsc{Cho$_{\rm r}$} where l stands
for left and r stands for right), the subprocess that has not been selected for execution is retained as an
initial subprocess in the target process of the generated transition. When both subprocesses are initial,
both rules for alternative composition are applicable, otherwise only one of them can be applied and in that
case it is the non-initial subprocess that can move, because the other one has been discarded at the moment
of the selection.

Every state corresponding to a non-final process has at least one outgoing transition, while every state
corresponding to a non-initial process has exactly one incoming transition due to the decoration of executed
actions. The labeled transition system underlying an initial process turns out to be a tree, whose branching
points correspond to occurrences of $+$.

	\begin{example}\label{ex:semantics}

The labeled transition systems generated by the rules in Table~\ref{tab:semantics} for the two initial
processes $a \, . \, \nil + a \, . \, \nil$ and $a \, . \, \nil$ are depicted below: \\[0.1cm]
\centerline{\includegraphics{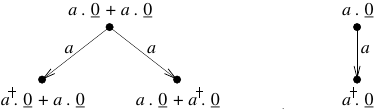}}
As far as the one on the left is concerned, we observe that, in the case of a standard process calculus, a
single $a$-transition from $a \, . \, \nil + a \, . \, \nil$ to $\nil$ would have been generated due to the
absence of action decorations within processes.
\fullbox

	\end{example}

%
\subsection{Strong Forward, Reverse, and Forward-Reverse Bisimilarities}
\label{sec:strong_bisim}
%

While forward bisimilarity considers only \emph{outgoing} transitions~\cite{Par81,Mil89a}, reverse
bisimilarity considers only \emph{incoming} transitions. Forward-reverse bisimilarity~\cite{PU07} considers
instead both outgoing transitions and incoming ones. Here are their \emph{strong} versions studied
in~\cite{BR23}, where strong means not abstracting from $\tau$-actions.

	\begin{definition}\label{def:forward_bisim}

We say that $P_{1}, P_{2} \in \procs$ are \emph{forward bisimilar}, written $P_{1} \sbis{\rm FB} P_{2}$, iff
$(P_{1}, P_{2}) \in \calb$ for some forward bisimulation $\calb$. A symmetric relation $\calb$ over~$\procs$
is a \emph{forward bisimulation} iff for all $(P_{1}, P_{2}) \in \calb$ and $a \in A$:

		\begin{itemize}

\item Whenever $P_{1} \arrow{a}{} P'_{1}$, then $P_{2} \arrow{a}{} P'_{2}$ with $(P'_{1}, P'_{2}) \in
\calb$.
\fullbox

		\end{itemize}

	\end{definition}

	\begin{definition}\label{def:reverse_bisim}

We say that $P_{1}, P_{2} \in \procs$ are \emph{reverse bisimilar}, written $P_{1} \sbis{\rm RB} P_{2}$, iff
$(P_{1}, P_{2}) \in \calb$ for some reverse bisimulation $\calb$. A symmetric relation $\calb$ over $\procs$
is a \emph{reverse bisimulation} iff for all $(P_{1}, P_{2}) \in \calb$ and $a \in A$:

		\begin{itemize}

\item Whenever $P'_{1} \arrow{a}{} P_{1}$, then $P'_{2} \arrow{a}{} P_{2}$ with $(P'_{1}, P'_{2}) \in
\calb$.
\fullbox

		\end{itemize}

	\end{definition}

	\begin{definition}\label{def:forward_reverse_bisim}

We say that $P_{1}, P_{2} \in \procs$ are \emph{forward-reverse bisimilar}, written $P_{1} \sbis{\rm FRB}
P_{2}$, iff $(P_{1}, P_{2}) \in \calb$ for some forward-reverse bisimulation $\calb$. A symmetric relation
$\calb$ over $\procs$ is a \emph{forward-reverse bisimulation} iff for all $(P_{1}, P_{2}) \in \calb$ and $a
\in A$:

		\begin{itemize}

\item Whenever $P_{1} \arrow{a}{} P'_{1}$, then $P_{2} \arrow{a}{} P'_{2}$ with $(P'_{1}, P'_{2}) \in
\calb$.

\item Whenever $P'_{1} \arrow{a}{} P_{1}$, then $P'_{2} \arrow{a}{} P_{2}$ with $(P'_{1}, P'_{2}) \in
\calb$.
\fullbox

		\end{itemize}

	\end{definition}

$\sbis{\rm FRB} \; \subsetneq \; \sbis{\rm FB} \cap \sbis{\rm RB}$ with the inclusion being strict because,
e.g., the two final processes $a^{\dag} . \, \nil$ and $a^{\dag} . \, \nil + c \, . \, \nil$ are identified
by $\sbis{\rm FB}$ (no outgoing transitions on both sides) and by $\sbis{\rm RB}$ (only an incoming
$a$-transition on both sides), but distinguished by $\sbis{\rm FRB}$ as in the latter process action $c$ is
enabled again after undoing $a$ (and hence there is an outgoing $c$-transition in addition to an outgoing
$a$-transition). Moreover, $\sbis{\rm FB}$ and $\sbis{\rm RB}$ are incomparable because for instance:
\cws{0}{\begin{array}{rcl}
a^{\dag} . \, \nil \sbis{\rm FB} \nil & \textrm{but} & a^{\dag} . \, \nil \not\sbis{\rm RB} \nil \\
a \, . \, \nil \sbis{\rm RB} \nil & \textrm{but} & a \, . \, \nil \not\sbis{\rm FB} \nil \\
\end{array}}
Note that that $\sbis{\rm FRB} \; = \; \sbis{\rm FB}$ over initial processes, with $\sbis{\rm RB}$ strictly
coarser, whilst $\sbis{\rm FRB} \; \neq \; \sbis{\rm RB}$ over final processes because, after going
backward, previously discarded subprocesses come into play again in the forward direction.

	\begin{example}\label{ex:bisim}

The two processes considered in Example~\ref{ex:semantics} are identified by all the three equivalences.
This is witnessed by any bisimulation that contains the pairs $(a \, . \, \nil + a \, . \, \nil, a \, . \,
\nil)$, $(a^{\dag} . \, \nil + a \, . \, \nil, a^{\dag} . \, \nil)$, and $(a \, . \, \nil + a^{\dag} . \,
\nil, a^{\dag} . \, \nil)$.
\fullbox

	\end{example}

As observed in~\cite{BR23}, it makes sense that $\sbis{\rm FB}$ identifies processes with a different past
and that $\sbis{\rm RB}$ identifies processes with a different future, in particular with $\nil$ that has
neither past nor future. However, for $\sbis{\rm FB}$ this breaks compositionality with respect to
alternative composition. As an example:
\cws{0}{\begin{array}{rcl}
a^{\dag} . \, b \, . \, \nil & \sbis{\rm FB} & b \, . \, \nil \\
a^{\dag} . \, b \, . \, \nil + c \, . \, \nil & \not\sbis{\rm FB} & b \, . \, \nil + c \, . \, \nil \\
\end{array}}
because in $a^{\dag} . \, b \, . \, \nil + c \, . \, \nil$ action $c$ is disabled due to the presence of the
already executed action $a^{\dag}$, while in $b \, . \, \nil + c \, . \, \nil$ action $c$ is enabled as
there are no past actions preventing it from occurring. Note that a similar phenomenon does not happen with
$\sbis{\rm RB}$ as $a^{\dag} . \, b \, . \, \nil \not\sbis{\rm RB} b \, . \, \nil$ due to the incoming
$a$-transition of~$a^{\dag} . \, b \, . \, \nil$.

This problem, which does not show up for $\sbis{\rm RB}$ and $\sbis{\rm FRB}$ because these two equivalences
cannot identify an initial process with a non-initial one, leads to the following variant of $\sbis{\rm FB}$
that is sensitive to the presence of the past.

	\begin{definition}\label{def:ps_forward_bisim}

We say that $P_{1}, P_{2} \in \procs$ are \emph{past-sensitive forward bisimilar}, written $P_{1} \sbis{\rm
FB:ps} P_{2}$, iff $(P_{1}, P_{2}) \in \calb$ for some past-sensitive forward bisimulation~$\calb$. A
relation $\calb$ over $\procs$ is a \emph{past-sensitive forward bisimulation} iff it is a forward
bisimulation such that $\initial(P_{1}) \Longleftrightarrow \initial(P_{2})$ for all $(P_{1}, P_{2}) \in
\calb$.
\fullbox

	\end{definition}

Now $\sbis{\rm FB:ps}$ is sensitive to the presence of the past:
\cws{0}{a^{\dag} . \, b \, . \, \nil \: \not\sbis{\rm FB:ps} \: b \, . \, \nil}
but can still identify non-initial processes having a different past:
\cws{0}{a_{1}^{\dag} \, . \, P \: \sbis{\rm FB:ps} \: a_{2}^{\dag} \, . \, P}
It holds that $\sbis{\rm FRB} \; \subsetneq \; \sbis{\rm FB:ps} \cap \sbis{\rm RB}$, with $\sbis{\rm FRB} \,
= \, \sbis{\rm FB:ps}$ over initial processes as well as $\sbis{\rm FB:ps}$ and $\sbis{\rm RB}$ being
incomparable because, e.g., for $a_{1} \neq a_{2}$:
\cws{5}{\begin{array}{rcl}
a_{1}^{\dag} \, . \, P \, \sbis{\rm FB:ps} \, a_{2}^{\dag} \, . \, P & \textrm{but} & a_{1}^{\dag} \, . \,
P \, \not\sbis{\rm RB} \, a_{2}^{\dag} \, . \, P \\
a_{1} \, . \, P \, \sbis{\rm RB} \, a_{2} \, . \, P & \textrm{but} & a_{1} \, . \, P \, \not\sbis{\rm FB:ps}
\, a_{2} \, . \, P \\
\end{array}}

In~\cite{BR23} it has been shown that all the considered strong bisimilarities are congruences with respect
to action prefix, while only $\sbis{\rm FB:ps}$, $\sbis{\rm RB}$, and $\sbis{\rm FRB}$ are congruences with
respect to alternative composition too, with $\sbis{\rm FB:ps}$ being the coarsest congruence with respect
to $+$ contained in $\sbis{\rm FB}$. Moreover, sound and complete equational characterizations have been
provided for the three congruences.

%
\subsection{Weak Forward, Reverse, and Forward-Reverse Bisimilarities}
\label{sec:weak_bisim}
%

In~\cite{BE23} \emph{weak} variants of forward, reverse, and forward-reverse bisimilarities have been
studied, which are capable of abstracting from $\tau$-actions. In the following definitions, $P
\warrow{\tau^{*}}{} P'$ means that $P' = P$ or there exists a nonempty sequence of finitely many
$\tau$-transitions such that the target of each of them coincides with the source of the subsequent one,
with the source of the first one being $P$ and the target of the last one being $P'$. Moreover,
$\warrow{\tau^{*}}{} \! \arrow{a}{} \! \warrow{\tau^{*}}{}$ stands for an $a$-transition possibly preceded
and followed by finitely many $\tau$-transitions. We further let $\bar{A} = A \setminus \{ \tau \}$.

	\begin{definition}\label{def:weak_forward_bisim}

We say that $P_{1}, P_{2} \in \procs$ are \emph{weakly forward bisimilar}, written $P_{1} \wbis{\rm FB}
P_{2}$, iff $(P_{1}, P_{2}) \in \calb$ for some weak forward bisimulation~$\calb$. A symmetric binary
relation $\calb$ over $\procs$ is a \emph{weak forward bisimulation} iff, whenever $(P_{1}, P_{2}) \in
\calb$, then:

		\begin{itemize}

\item Whenever $P_{1} \arrow{\tau}{} P'_{1}$, then $P_{2} \warrow{\tau^{*}}{} P'_{2}$ and $(P'_{1}, P'_{2})
\in \calb$.

\item Whenever $P_{1} \arrow{a}{} P'_{1}$ for $a \in \bar{A}$, then $P_{2} \warrow{\tau^{*}}{} \!
\arrow{a}{} \! \warrow{\tau^{*}}{} P'_{2}$ and $(P'_{1}, P'_{2}) \in \calb$.
\fullbox

		\end{itemize}

	\end{definition}

	\begin{definition}\label{def:weak_reverse_bisim}

We say that $P_{1}, P_{2} \in \procs$ are \emph{weakly reverse bisimilar}, written $P_{1} \wbis{\rm RB}
P_{2}$, iff $(P_{1}, P_{2}) \in \calb$ \linebreak for some weak reverse bisimulation~$\calb$. A symmetric
binary relation $\calb$ over $\procs$ is a \emph{weak reverse \linebreak bisimulation} iff, whenever
$(P_{1}, P_{2}) \in \calb$, then:

		\begin{itemize}

\item Whenever $P'_{1} \arrow{\tau}{} P_{1}$, then $P'_{2} \warrow{\tau^{*}}{} P_{2}$ and $(P'_{1}, P'_{2})
\in \calb$.

\item Whenever $P'_{1} \arrow{a}{} P_{1}$ for $a \in \bar{A}$, then $P'_{2} \warrow{\tau^{*}}{} \!
\arrow{a}{} \! \warrow{\tau^{*}}{} P_{2}$ and $(P'_{1}, P'_{2}) \in \calb$.
\fullbox

		\end{itemize}

	\end{definition}

	\begin{definition}\label{def:weak_forward_reverse_bisim}

We say that $P_{1}, P_{2} \in \procs$ are \emph{weakly forward-reverse bisimilar}, written $P_{1} \wbis{\rm
FRB} P_{2}$, iff $(P_{1}, P_{2}) \in \calb$ for some weak forward-reverse bisimulation~$\calb$. A symmetric
binary relation $\calb$ over $\procs$ is a \emph{weak forward-reverse bisimulation} iff, whenever $(P_{1},
P_{2}) \in \calb$, then:

		\begin{itemize}

\item Whenever $P_{1} \arrow{\tau}{} P'_{1}$, then $P_{2} \warrow{\tau^{*}}{} P'_{2}$ and $(P'_{1}, P'_{2})
\in \calb$.

\item Whenever $P_{1} \arrow{a}{} P'_{1}$ for $a \in \bar{A}$, then $P_{2} \warrow{\tau^{*}}{} \!
\arrow{a}{} \! \warrow{\tau^{*}}{} P'_{2}$ and $(P'_{1}, P'_{2}) \in \calb$.

\item Whenever $P'_{1} \arrow{\tau}{} P_{1}$, then $P'_{2} \warrow{\tau^{*}}{} P_{2}$ and $(P'_{1}, P'_{2})
\in \calb$.

\item Whenever $P'_{1} \arrow{a}{} P_{1}$ for $a \in \bar{A}$, then $P'_{2} \warrow{\tau^{*}}{} \!
\arrow{a}{} \! \warrow{\tau^{*}}{} P_{2}$ and $(P'_{1}, P'_{2}) \in \calb$.
\fullbox

		\end{itemize}

	\end{definition}

Each of the three weak bisimilarities is strictly coarser than the corresponding strong one. Similar to the
strong case, $\wbis{\rm FRB} \; \subsetneq \; \wbis{\rm FB} \cap \wbis{\rm RB}$ with $\wbis{\rm FB}$ and
$\wbis{\rm RB}$ being incomparable. Unlike the strong case, $\wbis{\rm FRB} \; \neq \; \wbis{\rm FB}$ over
initial processes. For instance, $\tau \, . \, a \, . \, \nil + a \, . \, \nil + b \, . \, \nil$ and $\tau
\, . \, a \, . \, \nil + b \, . \, \nil$ are identified by $\wbis{\rm FB}$ but told apart by $\wbis{\rm
FRB}$: if the former performs $a$, the latter responds with $\tau$ followed by~$a$ and if it subsequently
undoes $a$ thus becoming $\tau^{\dag} . \, a \, . \, \nil + b \, . \, \nil$ in which only $a$ is enabled,
the latter can only respond by undoing $a$ thus becoming $\tau \, . \, a \, . \, \nil + a \, . \, \nil + b
\, . \, \nil$ in which both $a$ and $b$ are enabled. An analogous counterexample with non-initial
$\tau$-actions is given by $c \, . \, (\tau \, . \, a \, . \, \nil + a \, . \, \nil + b \, . \, \nil)$ and
$c \, . \, (\tau \, . \, a \, . \, \nil + b \, . \, \nil)$.

As observed in~\cite{BE23}, $\wbis{\rm FB}$ suffers from the same compositionality problem with respect to
alternative composition as~$\sbis{\rm FB}$. Moreover, $\wbis{\rm FB}$ and $\wbis{\rm FRB}$ feature the same
compositionality problem as weak bisimilarity for standard forward-only processes~\cite{Mil89a}, i.e., for
$\wbis{} \, \in \{ \wbis{\rm FB}, \wbis{\rm FRB} \}$ it holds that:
\cws{0}{\begin{array}{rcl}
\tau \, . \, a \, . \, \nil & \: \wbis{} \: & a \, . \, \nil \\
\tau \, . \, a \, . \, \nil + b \, . \, \nil & \not\wbis{} & a \, . \, \nil + b \, . \, \nil \\
\end{array}}
because if $\tau \, . \, a \, . \, \nil + b \, . \, \nil$ performs $\tau$ thereby evolving to $\tau^{\dag} .
\, a \, . \, \nil + b \, . \, \nil$ where only $a$ is enabled in the forward direction, then $a \, . \, \nil
+ b \, . \, \nil$ can neither move nor idle in the attempt to evolve in such a way to match $\tau^{\dag} .
\, a \, . \, \nil + b \, . \, \nil$.

To solve both problems it is sufficient to redefine the two equivalences by making them sensitive to the
presence of the past, exactly as in the strong case for forward bisimilarity. By so doing, $\tau \, . \, a
\, . \, \nil$ is no longer identified with $a \, . \, \nil$: if the former performs $\tau$ thereby evolving
to $\tau^{\dag} . \, a \, . \, \nil$ and the latter idles, then $\tau^{\dag} . \, a \, . \, \nil$ and $a \,
. \, \nil$ are told apart because they are not both initial or non-initial.

	\begin{definition}\label{def:weak_ps_forward_bisim}

We say that $P_{1}, P_{2} \in \procs$ are \emph{weakly past-sensitive forward bisimilar}, written $P_{1}
\wbis{\rm FB:ps} P_{2}$, iff $(P_{1}, P_{2}) \in \calb$ for some weak past-sensitive forward bisimulation
$\calb$. A binary relation $\calb$ over $\procs$ is a \emph{weak past-sensitive forward bisimulation} iff it
is a weak forward bisimulation such that $\initial(P_{1}) \Longleftrightarrow \initial(P_{2})$ for all
$(P_{1}, P_{2}) \in \calb$.
\fullbox

	\end{definition}

	\begin{definition}\label{def:weak_ps_forward_reverse_bisim}

We say that $P_{1}, P_{2} \in \procs$ are \emph{weakly past-sensitive forward-reverse bisimilar}, written
\linebreak $P_{1} \wbis{\rm FRB:ps} P_{2}$, iff $(P_{1}, P_{2}) \in \calb$ for some weak past-sensitive
forward-reverse bisimulation $\calb$. A binary relation $\calb$ over $\procs$ is a \emph{weak past-sensitive
forward-reverse bisimulation} iff it is a weak forward-reverse bisimulation such that $\initial(P_{1})
\Longleftrightarrow \initial(P_{2})$ for all $(P_{1}, P_{2}) \in \calb$.
\fullbox

	\end{definition}

Like in the non-past-sensitive case, $\wbis{\rm FRB:ps} \; \neq \; \wbis{\rm FB:ps}$ over initial processes,
as shown by $\tau \, . \, a \, . \, \nil + a \, . \, \nil$ and $\tau \, . \, a \, . \, \nil$: if the former
performs $a$, the latter responds with $\tau$ followed by $a$ and if it subsequently undoes $a$ thus
becoming the non-initial process $\tau^{\dag} . \, a \, . \, \nil$, the latter can only respond by undoing
$a$ thus becoming the initial process $\tau \, . \, a \, . \, \nil + a \, . \, \nil$. An analogous
counterexample with non-initial $\tau$-actions is given again by $c \, . \, (\tau \, . \, a \, . \, \nil + a
\, . \, \nil + b \, . \, \nil)$ and $c \, . \, (\tau \, . \, a \, . \, \nil + b \, . \, \nil)$.

Observing that $\sbis{\rm FRB} \; \subsetneq \; \wbis{\rm FRB:ps}$ as the former naturally satisfies the
initiality condition, in~\cite{BE23} it has been shown that all the considered weak bisimilarities are
congruences with respect to action prefix, while only $\wbis{\rm FB:ps}$, $\wbis{\rm RB}$, and $\wbis{\rm
FRB:ps}$ are congruences with respect to alternative composition too, with $\wbis{\rm FB:ps}$ and $\wbis{\rm
FRB:ps}$ respectively being the coarsest congruences with respect to $+$ contained in $\wbis{\rm FB}$ and
$\wbis{\rm FRB}$. Sound and complete equational characterizations have been provided for the three
congruences.

%
%
\section{Modal Logic Characterizations}
\label{sec:modal_logics}
%
%

In this section we investigate modal logic characterizations for the three strong bisimilarities $\sbis{\rm
FB}$, $\sbis{\rm RB}$, and $\sbis{\rm FRB}$, the three weak bisimilarities $\wbis{\rm FB}$, $\wbis{\rm RB}$,
and $\wbis{\rm FRB}$, and the three past-sensitive variants $\sbis{\rm FB:ps}$, $\wbis{\rm FB:ps}$, and
$\wbis{\rm FRB:ps}$.

We start by introducing a general modal logic $\call$ from which we will take nine fragments to characterize
the nine aforementioned bisimilarities. It consists of Hennessy-Milner logic~\cite{HM85} extended with the
proposition init, the strong backward modality $\diam{a^{\dag}}{}$, the two weak forward modalities
$\wdiam{\tau}{}$ and $\wdiam{a}{}$, and the two weak backward modalities $\wdiam{\tau^{\dag}}{}$ and
$\wdiam{a^\dag}{}$ (where $a \in \bar{A}$ within weak modalities):
\cws{0}{\phi \:\: ::= \:\: \mathrm{true} \mid \mathrm{init} \mid \lnot\phi \mid \phi \land \phi \mid
\diam{a}{}\phi \mid \diam{a^\dag}{}\phi \mid \wdiam{\tau}{}\phi \mid \wdiam{a}{}\phi \mid
\wdiam{\tau^\dag}{}\phi \mid \wdiam{a^\dag}{}\phi}
The satisfaction relation $\models \: \subseteq \procs \times \call$ is defined by induction on the
syntactical structure of the formulas as follows:
\cws{6}{\begin{array}{rlll}
P & \models & \mathrm{true} & \text{for all } P \in \procs \\
P & \models & \mathrm{init} & \text{iff } \initial(P) \\
P & \models & \lnot\phi & \text{iff } P \not\models \phi \\
P & \models & \phi_1 \land \phi_2 & \text{iff } P \models_{\rm} \phi_1 \text{ and } P \models_{\rm} \phi_2
\\
P & \models & \diam{a}{}\phi & \text{iff there exists } P' \in \procs \text{ such that } P \arrow{a}{} P'
\text{ and } P' \models \phi \\
P & \models & \diam{a^\dag}{}\phi & \text{iff there exists } P' \in \procs \text{ such that } P' \arrow{a}{}
P \text{ and } P' \models \phi \\
P & \models & \wdiam{\tau}{}\phi & \text{iff there exists } P' \in \procs \text{ such that } P
\warrow{\tau^{*}}{} P' \text{ and } P' \models \phi \\
P & \models & \wdiam{a}{}\phi & \text{iff there exists } P' \in \procs \text{ such that } P
\warrow{\tau^{*}}{} \arrow{a}{} \warrow{\tau^{*}}{} P' \text{ and } P' \models \phi \\
P & \models & \wdiam{\tau^\dag}{}\phi & \text{iff there exists } P' \in \procs \text{ such that } P'
\warrow{\tau^{*}}{} P \text{ and } P' \models \phi \\
P & \models & \wdiam{a^\dag}{}\phi & \text{iff there exists } P' \in \procs \text{ such that } P'
\warrow{\tau^{*}}{} \arrow{a}{} \warrow{\tau^{*}}{} P \text{ and } P' \models \phi \\
\end{array}}

The use of backward operators is not new in the definition of properties of programs through temporal
logics~\cite{LPZ85} or modal logics~\cite{HS85}. In particular, in the latter work a logic with a past
operator was introduced to capture interesting properties of generalized labeled transition systems where
only visible actions are considered, in which setting it is proved that the equivalence induced by the
considered logic coincides with a generalization of the standard forward-only strong bisimilarity
of~\cite{Mil89a}. This result was later confirmed in~\cite{DV95} where it is shown that the addition of a
strong backward modality (interpreted over computation paths instead of states) provides no additional
discriminating power with respect to the Hennessy-Milner logic, i.e., the induced equivalence is again
strong bisimilarity.

In contrast, in our context -- in which all equivalences are defined over states -- the strong forward
bisimilarities $\sbis{\rm FB}$ and $\sbis{\rm FB:ps}$ do not coincide with the strong forward-reverse
bisimilarity $\sbis{\rm FRB}$ and this extends to their weak counterparts. In other words, the presence of
backward modalities matters. \linebreak It is worth noting that our two weak backward modalities are similar
to the ones considered in~\cite{DMV90,DV95} \linebreak to characterize weak back-and-forth bisimilarity
(defined over computation paths), which is finer than the standard forward-only weak bisimilarity
of~\cite{Mil89a} and coincides with branching bisimilarity~\cite{GW96}.

By taking suitable fragments of $\call$ we can characterize all the nine bisimilarities introduced in
Section~\ref{sec:background}. For each of the four strong bisimilarities $\sbis{B}$, where $B \in \{
\text{FB}, \text{FB:ps}, \text{RB}, \text{FRB} \}$, we can define the corresponding logic $\call_{B}$. The
same can be done for each of the five weak bisimilarities $\wbis{B}$, where $B \in \{ \text{FB},
\text{FB:ps}, \text{RB}, \text{FRB}, \text{FRB:ps} \}$, to obtain the corresponding logic
$\call_{B}^{\tau}$. All the considered fragments can be found in Table~\ref{tab:fragments}, which indicates
that the proposition $\mathrm{init}$ is needed only for the past-sensitive bisimilarities. The forthcoming
Theorems~\ref{thm:strong_logics} and~\ref{thm:weak_logics} show that each such fragment induces the intended
bisimilarity, in the sense that two processes are bisimilar iff they satisfy the same set of formulas of the
fragment at hand.

\pagebreak

	\begin{table}[t]

\[
\begin{array}{|l|c|c|c|c||c|c||c|c|c|c|}
\hline
& \mathrm{true} & \mathrm{init} & \lnot & \land & \diam{a}{} & \diam{a^\dag}{} & \wdiam{\tau}{} &
\wdiam{a}{} & \wdiam{\tau^\dag}{} & \wdiam{a^\dag}{} \\
\hline
\call_{\rm FB}^{} & \checkmark & & \checkmark & \checkmark & \checkmark & & & & & \\
\hline
\call_{\rm FB:ps}^{} & \checkmark & \checkmark & \checkmark & \checkmark & \checkmark & & & & & \\
\hline
\call_{\rm RB}^{} & \checkmark & & & & & \checkmark & & & & \\
\hline
\call_{\rm FRB} & \checkmark & & \checkmark & \checkmark & \checkmark & \checkmark & & & & \\
\hhline{|=|=|=|=|=||=|=||=|=|=|=|}
\call_{\rm FB}^{\tau} & \checkmark & & \checkmark & \checkmark & & & \checkmark & \checkmark & & \\
\hline
\call_{\rm FB:ps}^{\tau} & \checkmark & \checkmark & \checkmark & \checkmark & & & \checkmark & \checkmark &
& \\
\hline
\call_{\rm RB}^{\tau} & \checkmark & & & & & & & & \checkmark & \checkmark \\
\hline
\call_{\rm FRB}^{\tau} & \checkmark & & \checkmark & \checkmark & & & \checkmark & \checkmark & \checkmark &
\checkmark \\
\hline
\call_{\rm FRB:ps}^{\tau} & \checkmark & \checkmark & \checkmark & \checkmark & & & \checkmark & \checkmark
& \checkmark & \checkmark \\
\hline 
\end{array}
\]

\caption{Fragments of $\call$ characterizing the considered bisimilarities}
\label{tab:fragments}

	\end{table}

The technique used to prove the two theorems is inspired by the one employed in~\cite{AILS07} to show that
Hennessy-Milner logic characterizes the strong forward-only bisimilarity of~\cite{Mil89a}. The two
implications of either theorem are demonstrated separately. To prove that any pair of bisimilar processes
$P_1$ and $P_2$ satisfy the same formulas of the considered fragment, we assume that $P_1 \models \phi$ for
some formula $\phi$ and then we proceed by induction on the depth of $\phi$ to show that $P_2 \models \phi$
too, where the depth of a formula is defined by induction on the syntactical structure of the formula itself
as follows:
\cws{0}{\begin{array}{rcl}
\depth(\mathrm{true}) & = & 1 \\
\depth(\mathrm{init}) & = & 1 \\
\depth(\lnot\phi) & = & 1 + \depth(\phi) \\
\depth(\phi_1 \land \phi_2) & = & 1 + \mathrm{max}(\depth(\phi_1), \depth(\phi_2)) \\
\depth(\diam{a}{}\phi) & = & 1 + \depth(\phi) \\
\depth(\diam{a^\dag}{}\phi) & = & 1 + \depth(\phi) \\
\depth(\wdiam{\tau}{}\phi) & = & 1 + \depth(\phi) \\
\depth(\wdiam{a}{}\phi) & = & 1 + \depth(\phi) \\
\depth(\wdiam{\tau^\dag}{}\phi) & = & 1 + \depth(\phi) \\
\depth(\wdiam{a^\dag}{}\phi) & = & 1 + \depth(\phi) \\
\end{array}}
As for the reverse implication, we show that the relation $\calb$ formed by all pairs of processes $(P_1,
P_2)$ that satisfy the same formulas of the considered fragment is a bisimulation. More specifically,
starting from $(P_1, P_2) \in \calb$ we proceed by contradiction by assuming that, whenever $P_1$ has a move
to/from $P_1'$ with an action $a$, then there is no $P_2'$ such that $P_2$ has a move to/from $P_2'$ with
$a$ and $(P_1', P_2') \in \calb$. This entails that, for every $P_{2_{i}}$ forward/backward reachable from
$P_2$ by performing $a$, by definition of $\calb$ there exists some formula $\phi_i$ such that $P_1' \models
\phi_i$ and $P_{2_{i}}' \not\models \phi_i$, which leads to a formula with a forward/backward modality on
$a$ followed by $\bigwedge_i \phi_i$ that is satisfied by $P_1$ but not by~$P_2$, thereby contradicting
$(P_1, P_2) \in \calb$.

	\begin{theorem}\label{thm:strong_logics}

Let $P_1, P_2 \in \procs$ and $B \in \{ \text{FB}, \text{FB:ps}, \text{RB}, \text{FRB} \}$. Then $P_1
\sbis{B} P_2 \Longleftrightarrow \forall \phi \in \call_{B} \ldotp P_1 \models \phi \Leftrightarrow P_2
\models \phi$.
\fullbox

	\end{theorem}

	\begin{theorem}\label{thm:weak_logics}

Let $P_1, P_2 \in \procs$ and $B \in \{ \text{FB}, \text{FB:ps}, \text{RB}, \text{FRB}, \text{FRB:ps} \}$.
Then $P_1 \wbis{B} P_2 \Longleftrightarrow \forall \phi \in \call_{B}^{\tau} \ldotp \linebreak P_1 \models
\phi \Leftrightarrow P_2 \models \phi$.
\fullbox

	\end{theorem}

We conclude with the following observations:

	\begin{itemize}

\item The fragments that characterize the four forward bisimilarities $\sbis{\rm FB}$, $\sbis{\rm FB:ps}$,
$\wbis{\rm FB}$, and $\wbis{\rm FB:ps}$ are essentially identical to the Hennessy-Milner logic (first two
bisimilarities) and its weak variant (last two bisimilarities). The only difference is the possible presence
of the proposition $\mathrm{init}$, which is needed to distinguish between initial and non-initial
processes in the past-sensitive cases.

\item The fragments that characterize the two reverse bisimilarities $\sbis{\rm RB}$ and $\wbis{\rm RB}$
only include $\mathrm{true}$ and the backward modalities $\diam{a^\dag}{}$ (first bisimilarity) and
$\wdiam{\tau^\dag}{}$ and $\wdiam{a^\dag}{}$ (second bisimilarity). The absence of conjunction reflects the
fact that, when going backward, processes must follow exactly the sequence of actions they performed in the
forward direction and hence no choice is involved, consistent with every non-initial process having
precisely one incoming transition. In other words, the strong and weak reverse bisimilarities boil down to
strong and weak reverse trace equivalences, respectively, which consider traces obtained when going in the
backward direction.

\item The fragments that characterize the three forward-reverse bisimilarities $\sbis{\rm FRB}$, $\wbis{\rm
FRB}$, and $\wbis{\rm FRB:PS}$ are akin to the logic $\call_{\rm BF}$ introduced in~\cite{DMV90} to
characterize weak back-and-forth bisimilarity and branching bisimilarity. A crucial distinction between our
three fragments and $\call_{\rm BF}$ is that the former are interpreted over states while $\call_{\rm BF}$
is interpreted over computation paths. Moreover, as already mentioned, defining a strong variant of
$\call_{\rm BF}$ would yield a logic that characterizes strong bisimilarity, whereas in our setting
forward-only bisimilarities are different from forward-reverse ones and hence different logics are needed.

	\end{itemize}

%
%
\section{Weak Forward-Reverse Bisimilarity and Branching Bisimilarity}
\label{sec:branching_bisim}
%
%

In this section we establish a clear connection between weak forward-reverse bisimilarity and branching
bisimilarity~\cite{GW96}. Unlike the standard forward-only weak bisimilarity of~\cite{Mil89a}, branching
bisimilarity preserves the branching structure of processes even when abstracting from $\tau$-actions.

	\begin{definition}\label{def:branching_bisim}

We say that $P_{1}, P_{2} \in \procs$ are \emph{branching bisimilar}, written $P_{1} \wbis{\rm BB} P_{2}$,
iff $(P_{1}, P_{2}) \in \calb$ for some branching bisimulation~$\calb$. A symmetric binary relation $\calb$
over $\procs$ is a \emph{branching bisimulation} iff, whenever $(P_{1}, P_{2}) \in \calb$, then for all
$P_{1} \arrow{a}{} P'_{1}$ it holds that:

		\begin{itemize}

\item either $a = \tau$ and $(P'_{1}, P_{2}) \in \calb$;

\item or $P_{2} \warrow{\tau^{*}}{} \bar{P}_{2} \arrow{a}{} P'_{2}$ with $(P_{1}, \bar{P}_{2}) \in \calb$
and $(P'_{1}, P'_{2}) \in \calb$.
\fullbox

		\end{itemize}

	\end{definition}

Branching bisimilarity is known to have some relationships with reversibility. More precisely,
in~\cite{DMV90} strong and weak back-and-forth bisimilarities have been introduced over labeled transition
systems -- where outgoing transitions are considered in the forward bisimulation game while incoming
transitions are considered in the backward bisimulation game -- and respectively shown to coincide with the
standard forward-only strong bisimilarity of~\cite{Mil89a} and branching bisimilarity.

In the setting of~\cite{DMV90}, strong and weak back-and-forth bisimilarities have been defined over
computation paths rather than states so that, in the presence of concurrency, any backward computation is
\emph{constrained} to follow the same path as the corresponding forward computation, which is consistent
with an interleaving view of parallel composition. This is quite different from the forward-reverse
bisimilarity over states defined in~\cite{PU07}, which accounts for the fact that when going backward the
order in which independent transitions are undone may be different from the order in which they were
executed in the forward direction, thus leading to a truly concurrent semantics.

Since in our setting we consider only sequential processes, hence any backward computation \emph{naturally}
follows the same path as the corresponding forward computation, we are neutral with respect to interleaving
vs.\ true concurrency. Like in~\cite{DMV90} we define a single transition relation and then we distinguish
between outgoing transitions and incoming transitions in the bisimulation game. However,
unlike~\cite{DMV90}, our bisimilarities are defined over states as in~\cite{Mil89a,GW96,PU07}, not over
paths. In the rest of this section we show that our weak forward-reverse bisimilarity \emph{over states}
coincides with branching bisimilarity \linebreak by following the proof strategy adopted in~\cite{DMV90} for
weak back-and-forth bisimilarity.

First of all, we prove that, like branching bisimilarity, our weak forward-reverse bisimilarity satisfies
the \emph{stuttering property}~\cite{GW96}. This means that, given a sequence of finitely many
$\tau$-transitions, if the source process of the first transition and the target process of the last
transition are equivalent to each other, then all the intermediate processes are equivalent to them too --
see $P_{2} \warrow{\tau^{*}}{} \bar{P}_{2}$ in Definition~\ref{def:branching_bisim} when $P_{1}, P_{2},
\bar{P}_{2}$ are pairwise related by the maximal branching bisimulation $\wbis{\rm BB}$. In other words,
while traversing the considered sequence of $\tau$-transitions, we remain in the same equivalence class of
processes, not only in the forward direction but -- as we are talking about weak forward-reverse
bisimilarity -- \emph{also in the backward direction}. This property does not hold in the case of the
standard forward-only weak bisimilarity of~\cite{Mil89a}.

	\begin{lemma}\label{lem:stuttering_prop}

Let $n \in \natns_{> 0}$, $P_{i} \in \procs$ for all $0 \le i \le n$, and $P_{i} \arrow{\tau}{} P_{i + 1}$
for all $0 \le i \le n - 1$. If $P_{0} \wbis{\rm FRB} P_{n}$ then $P_{i} \wbis{\rm FRB} P_{0}$ for all $0
\le i \le n$.

		\begin{proof}
Consider the reflexive and symmetric binary relation $\calb = \cup_{i \in \natns} \calb_{i}$ over $\procs$
where:

			\begin{itemize}

\item $\calb_{0} = \:\: \wbis{\rm FRB}$.

\item $\calb_{i} = \calb_{i - 1} \cup \{ (P, P'), (P', P) \in \procs \times \procs \mid \exists P'' \in
\procs \ldotp (P, P'') \in \calb_{i - 1} \land P \warrow{\tau^{*}}{} P' \arrow{\tau}{} P'' \}$ for all $i
\in \natns_{> 0}$.

			\end{itemize}

\noindent
We start by proving that $\calb$ satisfies the stuttering property, i.e.,
given $n \in \natns_{> 0}$ and $P_{i} \in \procs$ for all $0 \le i \le n$, if $P_{i} \arrow{\tau}{} P_{i +
1}$ for all $0 \le i \le n - 1$ and $(P_{0}, P_{n}) \in \calb$, then $(P_{i}, P_{0}) \in \calb$ for all $0
\le i \le n$. We proceed by induction on $n$:

			\begin{itemize}

\item If $n = 1$ then the considered computation is simply $P_{0} \arrow{\tau}{} P_{1}$ with $(P_{0}, P_{1})
\in \calb$ and hence trivially $(P_{i}, P_{0}) \in \calb$ for all $0 \le i \le 1$ as $\calb$ is reflexive --
$(P_{0}, P_{0}) \in \calb$ -- and symmetric -- $(P_{1}, P_{0}) \in \calb$.

\item Let $n > 1$. Since $(P_{0}, P_{n}) \in \calb$, there must exist $m \in \natns$ such that $(P_{0},
P_{n}) \in \calb_{m}$. Let us consider the smallest such $m$. Then $(P_{0}, P_{n - 1}) \in \calb_{m + 1}$ by
definition of $\calb_{m + 1}$, hence $(P_{0}, P_{n - 1}) \in \calb$. From the induction hypothesis it
follows that $(P_{i}, P_{0}) \in \calb$ for all $0 \le i \le n - 1$, hence $(P_{i}, P_{0}) \in \calb$ for
all $0 \le i \le n$ because $(P_{0}, P_{n}) \in \calb$ and $\calb$ is symmetric so that $(P_{n}, P_{0}) \in
\calb$.

			\end{itemize}

\noindent
We now prove that every symmetric relation $\calb_{i}$ is a weak forward-reverse bisimulation. We proceed by
induction on $i \in \natns$:

			\begin{itemize}

\item If $i = 0$ then $\calb_{i}$ is the maximal weak forward-reverse bisimulation.

\item Let $i \ge 1$ and suppose that $\calb_{i - 1}$ is a weak forward-reverse bisimulation. Given $(P, P')
\in \calb_{i}$, assume that $P \arrow{a}{} Q$ (resp.\ $Q \arrow{a}{} P$) where $a \in A$. There are two
cases:

				\begin{itemize}

\item If $(P, P') \in \calb_{i - 1}$ then by the induction hypothesis $a = \tau$ and $P' \warrow{\tau^{*}}{}
Q'$ (resp.\ $Q' \warrow{\tau^{*}}{} P'$) \linebreak or $a \neq \tau$ and $P' \warrow{\tau^{*}}{} \arrow{a}{}
\warrow{\tau^{*}}{} Q'$ (resp.\ $Q' \warrow{\tau^{*}}{} \arrow{a}{} \warrow{\tau^{*}}{} P'$) with $(Q, Q')
\in \calb_{i - 1}$ and hence $(Q, Q') \in \calb_{i}$ as $\calb_{i - 1} \subseteq \calb_{i}$ by definition of
$\calb_{i}$.

\item If instead $(P, P') \notin \calb_{i - 1}$ then from $(P, P') \in \calb_{i}$ it follows that $\exists
P'' \in \procs \ldotp (P, P'') \in \calb_{i - 1} \land P \warrow{\tau^{*}}{} P' \arrow{\tau}{} P''$. There
are two subcases:

					\begin{itemize}

\item In the forward case, i.e., $P \arrow{a}{} Q$, there are two further subcases:

						\begin{itemize}

\item If $(Q, P'') \in \calb_{i - 1}$ and $a = \tau$, then from $P' \arrow{\tau}{} P''$ it follows that $P'
\warrow{\tau^{*}}{} P''$ with $(Q, P'') \in \calb_{i}$ as $\calb_{i - 1} \subseteq \calb_{i}$.

\item Otherwise from $(P, P'') \in \calb_{i - 1}$ and the induction hypothesis it follows that \linebreak
$P'' \warrow{\tau^{*}}{} \arrow{a}{} \warrow{\tau^{*}}{} P'''$ with $(Q, P''') \in \calb_{i - 1}$ so that
$P' \warrow{\tau^{*}}{} \arrow{a}{} \warrow{\tau^{*}}{} P'''$ with $(Q, P''') \in \calb_{i}$ as $\calb_{i -
1} \subseteq \calb_{i}$.

						\end{itemize}

\item In the backward case, i.e., $Q \arrow{a}{} P$, it suffices to note that from $P \warrow{\tau^{*}}{}
P'$ it follows that $Q \arrow{a}{} \warrow{\tau^{*}}{} P'$.

					\end{itemize}

				\end{itemize}

			\end{itemize}

\noindent
Since $\calb$ is the union of countably many weak forward-reverse bisimulations, it holds that $\calb
\subseteq \:\: \wbis{\rm FRB}$. On the other hand, $\wbis{\rm FRB} \:\: \subseteq \calb$ by definition of
$\calb_{0}$. In conclusion $\calb = \:\: \wbis{\rm FRB}$ -- i.e., no relation $\calb_{i}$ for $i \in
\natns_{> 0}$ adds further pairs with respect to $\calb_{0}$ -- and hence $\wbis{\rm FRB}$ satisfies the
stuttering property because so does $\calb$.
\fullbox

		\end{proof}

	\end{lemma}

Note that the lemma above considers $\wbis{\rm FRB}$, not $\wbis{\rm FRB:PS}$. Indeed the stuttering
property does not hold for $\wbis{\rm FRB:PS}$ when $\initial(P_{0})$, because in that case a $\tau$-action
would be decorated inside $P_{1}$ and hence $P_{1} \not\wbis{\rm FRB:ps} P_{0}$. Therefore $\wbis{\rm
FRB:PS}$ satisfies the stuttering property only over non-initial processes.

Secondly, we prove that $\wbis{\rm FRB}$ satisfies the \emph{cross property}~\cite{DMV90}. This means that,
whenever two processes reachable from two $\wbis{\rm FRB}$-equivalent processes can perform a sequence of
finitely many \linebreak $\tau$-transitions such that each of the two target processes is $\wbis{\rm
FRB}$-equivalent to the source process of the other sequence, then the two target processes are $\wbis{\rm
FRB}$-equivalent to each other as well.

	\begin{lemma}\label{lem:cross_prop}

Let $P_{1}, P_{2} \in \procs$ be such that $P_{1} \wbis{\rm FRB} P_{2}$. For all $P'_{1}, P''_{1} \in
\procs$ reachable from $P_{1}$ such that $P'_{1} \warrow{\tau^{*}}{} P''_{1}$ and for all $P'_{2}, P''_{2}
\in \procs$ reachable from $P_{2}$ such that $P'_{2} \warrow{\tau^{*}}{} P''_{2}$, if $P'_{1} \wbis{\rm FRB}
P''_{2}$ and $P''_{1} \wbis{\rm FRB} P'_{2}$ then $P''_{1} \wbis{\rm FRB} P''_{2}$.

%
%
%
%
%

		\begin{proof}
Given $P_{1}, P_{2} \in \procs$ with $P_{1} \wbis{\rm FRB} P_{2}$, consider the symmetric relation $\calb =
\:\: \wbis{\rm FRB} \! \cup \: \{ (P''_{1}, P''_{2}), (P''_{2}, P''_{1}) \linebreak \in \procs \times \procs
\mid \exists P'_{1}, P'_{2} \in \procs \textrm{ resp.\ reachable from $P_{1}, P_{2}$} \ldotp \, P'_{1}
\warrow{\tau^{*}}{} P''_{1} \land P'_{2} \warrow{\tau^{*}}{} P''_{2} \land P'_{1} \wbis{\rm FRB} P''_{2}
\land P''_{1} \wbis{\rm FRB} P'_{2} \}$. \linebreak The result follows by proving that $\calb$ is a weak
forward-reverse bisimulation, because this implies that $P''_{1} \wbis{\rm FRB} P''_{2}$ for every
additional pair -- i.e., $\calb$ satisfies the cross property -- as well as $\calb = \:\: \wbis{\rm FRB}$ --
hence $\wbis{\rm FRB}$ satisfies the cross property too. \\
Let $(P''_{1}, P''_{2}) \in \calb \, \setminus \! \wbis{\rm FRB}$ to avoid trivial cases. Then there exist
$P'_{1}, P'_{2} \in \procs$ respectively reachable from $P_{1}, P_{2}$ such that $P'_{1} \warrow{\tau^{*}}{}
P''_{1}$, $P'_{2} \warrow{\tau^{*}}{} P''_{2}$, $P'_{1} \wbis{\rm FRB} P''_{2}$, and $P''_{1} \wbis{\rm FRB}
P'_{2}$. There are two cases:

			\begin{itemize}

\item In the forward case, assume that $P''_{1} \arrow{a}{} P'''_{1}$, from which it follows that $P'_{1}
\warrow{\tau^{*}}{} P''_{1} \arrow{a}{} P'''_{1}$. Since $P'_{1} \wbis{\rm FRB} P''_{2}$, we obtain $P''_{2}
\warrow{\tau^{*}}{} \arrow{a}{} \warrow{\tau^{*}}{} P'''_{2}$, or $P''_{2} \warrow{\tau^{*}}{} P'''_{2}$
when $a = \tau$, with $P'''_{1} \wbis{\rm FRB} P'''_{2}$ and hence $(P'''_{1}, P'''_{2}) \in \calb$.
Starting from $P''_{2} \arrow{a}{} P'''_{2}$ one exploits $P'_{2} \warrow{\tau^{*}}{} P''_{2}$ and $P''_{1}
\wbis{\rm FRB} P'_{2}$ instead.

\item In the backward case, assume that $P'''_{1} \arrow{a}{} P''_{1}$. Since $P''_{1} \wbis{\rm FRB}
P'_{2}$, we obtain $P'''_{2} \warrow{\tau^{*}}{} \arrow{a}{} \warrow{\tau^{*}}{} P'_{2}$, so that $P'''_{2}
\warrow{\tau^{*}}{} \arrow{a}{} \warrow{\tau^{*}}{} P''_{2}$, or $P'''_{2} \warrow{\tau^{*}}{} P'_{2}$ when
$a = \tau$, so that $P'''_{2} \warrow{\tau^{*}}{} P''_{2}$, with $P'''_{1} \wbis{\rm FRB} P'''_{2}$ and
hence $(P'''_{1}, P'''_{2}) \in \calb$. Starting from $P'''_{2} \arrow{a}{} P''_{2}$ one exploits $P'_{1}
\wbis{\rm FRB} P''_{2}$ and $P'_{1} \warrow{\tau^{*}}{} P''_{1}$ instead.
\fullbox

			\end{itemize}

		\end{proof}

	\end{lemma}

We are now in a position of proving that $\wbis{\rm FRB}$ coincides with $\wbis{\rm BB}$. This only holds
over initial processes though. As an example, $a_{1}^{\dag} . \, b \, . \, P \wbis{\rm BB} a_{2}^{\dag} . \,
b \, . \, P$ but $a_{1}^{\dag} . \, b \, . \, P \not\wbis{\rm FRB} a_{2}^{\dag} . \, b \, . \, P$ when
$a_{1} \neq a_{2}$.

	\begin{theorem}\label{thm:wfrb_bb}

Let $P_{1}, P_{2} \in \procs$ be initial. Then $P_{1} \wbis{\rm FRB} P_{2}$ iff $P_{1} \wbis{\rm BB} P_{2}$.

		\begin{proof}
Given two initial processes $P_{1}, P_{2} \in \procs$, we divide the proof into two parts:

			\begin{itemize}

\item Given a weak forward-reverse bisimulation $\calb$ witnessing $P_{1} \wbis{\rm FRB} P_{2}$ and only
containing all the pairs of $\wbis{\rm FRB}$-equivalent processes reachable from $P_{1}$ and $P_{2}$ so that
Lemma~\ref{lem:cross_prop} is applicable to~$\calb$, we prove that $\calb$ is a branching bisimulation too.
Let $(Q_{1}, Q_{2}) \in \calb$, where $Q_{1}$ is reachable from~$P_{1}$ while $Q_{2}$ is reachable from
$P_{2}$, and assume that $Q_{1} \arrow{a}{} Q'_{1}$. There are two cases:

				\begin{itemize}

\item Suppose that $a = \tau$ and $Q_{2} \warrow{\tau^{*}}{} Q'_{2}$ with $(Q'_{1}, Q'_{2}) \in \calb$. This
means that we have a sequence of $n \ge 0$ transitions of the form $Q_{2, i} \arrow{\tau}{} Q_{2, i + 1}$
for all $0 \le i \le n - 1$ where $Q_{2, 0}$ is $Q_{2}$ while $Q_{2, n}$ is $Q'_{2}$ so that $(Q'_{1}, Q_{2,
n}) \in \calb$. \\
If $n = 0$ then $Q'_{2}$ is $Q_{2}$ and we are done because $(Q'_{1}, Q_{2}) \in \calb$, otherwise from
$Q_{2, n}$ we go back to $Q_{2, n - 1}$ via $Q_{2, n - 1} \arrow{\tau}{} Q_{2, n}$. If $Q'_{1}$ stays idle
so that $(Q'_{1}, Q_{2, n - 1}) \in \calb$ and $n = 1$ then we are done because $(Q'_{1}, Q_{2}) \in \calb$,
otherwise we go back to $Q_{2, n - 2}$ via $Q_{2, n - 2} \arrow{\tau}{} Q_{2, n - 1}$. \linebreak By
repeating this procedure, either we get to $(Q'_{1}, Q_{2, 0}) \in \calb$ and we are done because $(Q'_{1},
Q_{2}) \in \calb$, or for some $0 < m \le n$ such that $(Q'_{1}, Q_{2, m}) \in \calb$ we have that the
incoming transition $Q_{2, m - 1} \arrow{\tau}{} Q_{2, m}$ is matched by $\bar{Q}_{1} \warrow{\tau^{*}}{}
Q_{1} \arrow{\tau}{} Q'_{1}$ with $(\bar{Q}_{1}, Q_{2, m - 1}) \in \calb$. \linebreak In the latter case,
since $\bar{Q}_{1} \warrow{\tau^{*}}{} Q_{1}$, $Q_{2} \warrow{\tau^{*}}{} Q_{2, m - 1}$, $(\bar{Q}_{1},
Q_{2, m - 1}) \in \calb$, and $(Q_{1}, Q_{2}) \in \calb$, from Lemma~\ref{lem:cross_prop} it follows that
$(Q_{1}, Q_{2, m - 1}) \in \calb$.
In conclusion $Q_{2} \warrow{\tau^{*}}{} Q_{2, m - 1} \arrow{\tau}{} Q_{2, m}$ with $(Q_{1}, Q_{2, m - 1})
\in \calb$ and $(Q'_{1}, Q_{2, m}) \in \calb$.

\item Suppose that $a \neq \tau$ and $Q_{2} \warrow{\tau^{*}}{} \bar{Q}_{2} \arrow{a}{} \bar{Q}'_{2}
\warrow{\tau^{*}}{} Q'_{2}$ with $(Q'_{1}, Q'_{2}) \in \calb$. \\
From $\bar{Q}'_{2} \warrow{\tau^{*}}{} Q'_{2}$ and $(Q'_{1}, Q'_{2}) \in \calb$ it follows that
$\bar{Q}'_{1} \warrow{\tau^{*}}{} Q'_{1}$ with $(\bar{Q}'_{1}, \bar{Q}'_{2}) \in \calb$. Since $Q'_{1}$
already has an incoming $a$-transition from $Q_{1}$ and every non-initial process has exactly one incoming
transition, we derive that $\bar{Q}'_{1}$ is $Q'_{1}$ and hence $(Q'_{1}, \bar{Q}'_{2}) \in \calb$. \\
From $\bar{Q}_{2} \arrow{a}{} \bar{Q}'_{2}$ and $(Q'_{1}, \bar{Q}'_{2}) \in \calb$ it follows that
$\bar{Q}_{1} \warrow{\tau^{*}}{} Q_{1} \arrow{a}{} Q'_{1}$ with $(\bar{Q}_{1}, \bar{Q}_{2}) \in \calb$.
Since $\bar{Q}_{1} \warrow{\tau^{*}}{} Q_{1}$, $Q_{2} \warrow{\tau^{*}}{} \bar{Q}_{2}$, $(\bar{Q}_{1},
\bar{Q}_{2}) \in \calb$, and $(Q_{1}, Q_{2}) \in \calb$, from Lemma~\ref{lem:cross_prop} it follows that
$(Q_{1}, \bar{Q}_{2}) \in \calb$. \\
In conclusion $Q_{2} \warrow{\tau^{*}}{} \bar{Q}_{2} \arrow{a}{} \bar{Q}'_{2}$ with $(Q_{1}, \bar{Q}_{2})
\in \calb$ and $(Q'_{1}, \bar{Q}'_{2}) \in \calb$.

				\end{itemize}

\item Given a branching bisimulation $\calb$ witnessing $P_{1} \wbis{\rm BB} P_{2}$ and only containing all
the processes reachable from $P_{1}$ and $P_{2}$, we prove that $\calb$ is a weak forward-reverse
bisimulation too. \linebreak Let $(Q_{1}, Q_{2}) \in \calb$ with $Q_{1}$ reachable from $P_{1}$ and $Q_{2}$
reachable from $P_{2}$. There are two cases:

				\begin{itemize}

\item In the forward case, assume that $Q_{1} \arrow{a}{} Q'_{1}$. Then either $a = \tau$ and $(Q'_{1},
Q_{2}) \in \calb$, hence $Q_{2} \warrow{\tau^{*}}{} Q_{2}$ with $(Q'_{1}, Q_{2}) \in \calb$, or $Q_{2}
\warrow{\tau^{*}}{} \bar{Q}_{2} \arrow{a}{} Q'_{2}$ with $(Q_{1}, \bar{Q}_{2}) \in \calb$ and $(Q'_{1},
Q'_{2}) \in \calb$, hence $Q_{2} \warrow{\tau^{*}}{} \arrow{a}{} \warrow{\tau^{*}}{} Q'_{2}$ with $(Q'_{1},
Q'_{2}) \in \calb$.

\item In the backward case -- which cannot be the one of $(P_{1}, P_{2}) \in \calb$ as both processes are
initial -- assume that $Q'_{1} \arrow{a}{} Q_{1}$. There are two subcases:

					\begin{itemize}

\item Suppose that $Q'_{1}$ is $P_{1}$. Then either $a = \tau$ and $(Q'_{1}, Q_{2}) \in \calb$, where
$Q_{2}$ is $P_{2}$ and $Q_{2} \warrow{\tau^{*}}{} Q_{2}$, or $Q'_{2} \warrow{\tau^{*}}{} \bar{Q}_{2}
\arrow{a}{} Q_{2}$ with $(Q'_{1}, \bar{Q}_{2}) \in \calb$ and $(Q'_{1}, Q'_{2}) \in \calb$, where $Q'_{2}$
is~$P_{2}$ and $Q'_{2} \warrow{\tau^{*}}{} \arrow{a}{} \warrow{\tau^{*}}{} Q_{2}$.

\item If $Q'_{1}$ is not $P_{1}$, then $P_{1}$ reaches $Q'_{1}$ with a sequence of moves that are
$\calb$-compatible with those with which $P_{2}$ reaches some $Q'_{2}$ such that $(Q'_{1}, Q'_{2}) \in
\calb$ as $\calb$ only contains all the processes reachable from $P_{1}$ and $P_{2}$. Therefore either $a =
\tau$ and $(Q_{1}, Q'_{2}) \in \calb$, where $Q'_{2}$ is $Q_{2}$ and $Q_{2} \warrow{\tau^{*}}{} Q_{2}$, or
$Q'_{2} \warrow{\tau^{*}}{} \bar{Q}_{2} \arrow{a}{} Q_{2}$ with $(Q'_{1}, \bar{Q}_{2}) \in \calb$ in
addition to $(Q'_{1}, Q'_{2}) \in \calb$ and $(Q_{1}, Q_{2}) \in \calb$, where $Q'_{2} \warrow{\tau^{*}}{}
\arrow{a}{} \warrow{\tau^{*}}{} Q_{2}$.
\fullbox

					\end{itemize}

				\end{itemize}

			\end{itemize}

		\end{proof}

	\end{theorem}

According to the logical characterizations of branching bisimilarity shown in~\cite{DV95}, this result opens
the way to further logical characterizations of $\wbis{\rm FRB}$ over initial processes in addition to the
one of Section~\ref{sec:modal_logics} based on forward and backward modalities:

	\begin{itemize}

\item The first additional characterization replaces the two aforementioned modalities with an until
operator $\phi_{1} \wdiam{a}{} \phi_{2}$. This is satisfied by a process $P$ iff either $a = \tau$ with $P$
satisfying $\phi_{2}$, or $P \warrow{\tau^{*}}{} \bar{P} \arrow{a}{} P'$ with every process along $P
\warrow{\tau^{*}}{} \bar{P}$ satisfying $\phi_{1}$ and $P'$ satisfying $\phi_{2}$.

\item The second additional characterization is given by the temporal logic CTL$^{*}$ without the next
operator, thanks to a revisitation of the stuttering equivalence of~\cite{BCG88} and the bridge between
Kripke structures (in which states are labeled with propositions) and labeled transition systems (in which
transitions are labeled with actions) built in~\cite{DV95}.

	\end{itemize}

%
%
\section{Conclusion}
\label{sec:concl}
%
%

In this paper we have investigated modal logic characterizations of forward, reverse, and forward-reverse
bisimilarities, both strong and weak, over nondeterministic reversible sequential processes. While previous
work~\cite{BR23,BE23} has addressed compositionality and axiomatizations of those bisimilarities, here the
focus has been on identifying suitable modal logics, which are essentially variants of the Hennessy-Milner
logic~\cite{HM85}, such that two processes are bisimilar iff they satisfy the same set of formulas of the
corresponding modal logic.

The additional backward modalities used in this paper are inspired by those in~\cite{DMV90}, with the
important difference that bisimilarities and modal interpretations in the former are defined over states --
as is usual -- while those in the latter are defined over computation paths. The modal logic
characterizations have revealed that strong and weak reverse bisimilarities respectively boil down to strong
and weak reverse trace equivalences. Moreover, we have shown that weak forward-reverse bisimilarity
coincides with branching bisimilarity~\cite{GW96} over initial processes, thus providing two further logical
characterizations for the former thanks to~\cite{DV95}.

The study carried out in this paper can contribute, together with the results in~\cite{BR23,BE23}, to the
development of a fully-fledged process algebraic theory of reversible systems. On a more applicative side,
following~\cite{Cle90} we also observe that the established modal logic characterizations are useful to
provide diagnostic information because, whenever two processes are not bisimilar, then there exists at least
one formula in the modal logic corresponding to the considered bisimilarity that is satisfied by only one of
the two processes and hence can explain the inequivalence.

\medskip
\noindent
\textbf{Acknowledgments.}
This research has been supported by the PRIN project \emph{NiRvAna -- Noninterference and Reversibility
Analysis in Private Blockchains}.

\bibliographystyle{eptcs}
\bibliography{biblio}

\end{document}